\documentclass[conference,compsoc]{IEEEtran}
\usepackage{comment}
\usepackage{color}
\usepackage{cite}
\ifCLASSINFOpdf
   \usepackage[pdftex]{graphicx}
   \usepackage{subfigure}
\else
\fi

\usepackage{booktabs}
\usepackage{graphicx}
\usepackage[graphicx]{realboxes}
\usepackage{float}

\begin{document}

\title{Time Series Analysis and Correlation of Subway Turnstile Usage and COVID-19 Prevalence in New York City}

\author{\IEEEauthorblockN{Sina Fathi-Kazerooni\IEEEauthorrefmark{1}, Roberto Rojas-Cessa\IEEEauthorrefmark{1}, Ziqian Dong\IEEEauthorrefmark{2}, and Vatcharapan Umpaichitra\IEEEauthorrefmark{3}}
\IEEEauthorblockA{\IEEEauthorrefmark{1}Department of Electrical and Computer Engineering\\Newark College of Engineering \\ New Jersey Institute of Technology, Newark, NJ 07102.\\
\IEEEauthorrefmark{2}Department of Electrical and Computer Engineering\\ College of Engineering and Computing Sciences\\ New York Institute of Technology, New York, NY 10023.\\
\IEEEauthorrefmark{3}Department of Pediatrics \\ SUNY Downstate Health Sciences University, Brooklyn, NY 11203.}
  Email: \{sina.fathi.kazerooni, Roberto.Rojas-cessa\}@njit.edu, ziqian.dong@nyit.edu, Vatcharapan.Umpaichitra@downstate.edu}

\maketitle

\thispagestyle{plain}

\begin{abstract}
In this paper, we show a strong correlation between turnstile usage data of the New York City subway provided by the Metropolitan Transport Authority of New York City and COVID-19 deaths and cases reported by the New York City Department of Health. The turnstile usage data not only indicate the usage of the city's subway but also people's activity that promoted the large prevalence of COVID-19 city dwellers experienced from March to May of 2020. While this correlation is apparent, no proof has been provided before. Here we demonstrate this correlation through the application of a long short-term memory neural network. We show that the correlation of COVID-19 prevalence and deaths considers the incubation and symptomatic phases on reported deaths. Having established this correlation, we estimate the dates when the number of COVID-19 deaths and cases would approach zero after the reported number of deaths were decreasing by using the Auto-Regressive Integrated Moving Average model. We also estimate the dates when the first cases and deaths occurred by back-tracing the data sets and compare them to the reported dates.
\end{abstract}


%
\IEEEpeerreviewmaketitle

\section{Introduction}


New York City (NYC) has been the locus of a major prevalence of coronavirus disease 2019 (COVID-19), caused by severe acute respiratory syndrome coronavirus 2 (SARS-CoV-2), in early 2020 \cite{nyc_epicenter}. Many people have been hospitalized and also many have lost their lives to this disease. It is not clear when the first cases occurred in the city but the numbers of cases and deaths in the city peaked in mid April 2020 \cite{nyc_COVID_data}. This pandemic has also brought and continues to bring economical hardship to city dwellers and people worldwide, especially those in the lowest economical strata \cite{wadhera2020variation}. Measures taken to contain the spreading of SARS-CoV-2 including social distancing, business shut-downs, and shelter-in-place have significant impact on the economy \cite{nicola2020socio}, as up to the time of writing this paper no vaccine or other contagion control method has been developed. Essential workers often have higher risks of exposure to the virus because of the lack of options of working from home.

As city dwellers mainly rely on public transportation, and more notably the subway, to move around, it is expected that a highly contagious virus such as SARS-CoV-2 would easily spread through the fabric of NYC population.   
It is expected that a crowded NYC subway, as an enclosed environment and a major indicator of people's mobility, would be directly correlated with spreading of SARS-CoV-2.

While the NYC subway was expected to be a major vehicle for the transmission of COVID-19 in NYC in early 2020 \cite{harris2020subways}, the correlation between the subway ridership and COVID-19 prevalence and deaths has not been presented until now. 
In this paper, we show this correlation on the public turnstile usage data of NYC subway provided by the NYC Metropolitan Transportation Authority (MTA) \cite{MTA} and the statistics of the confirmed COVID-19 cases and deaths provided by the NYC Department of Health (DOH) \cite{NYChealth}. 

There is a large number of prediction models for the COVID-19 pandemic \cite{KUCHARSKI2020553,gosce2018analysing,mo2020modeling,ribeiro2020short,zhang2020estimation,li2020propagation,fanelli2020analysis,ceylan2020estimation,tosepu2020correlation,10.3389/fphy.2020.00127,qin2020prediction,al2020optimization,kuniya2020prediction,anastassopoulou2020data,petropoulos2020forecasting,yang2020modified,chakraborty2020real}. Much of the literature focuses on modeling the prevalence of COVID-19 in different countries and partially enclosed environments, such as cruise ships \cite{KUCHARSKI2020553,ribeiro2020short, zhang2020estimation,li2020propagation,fanelli2020analysis,ceylan2020estimation,tosepu2020correlation}. These models use local and regional data to examine the correlation of COVID-19 cases with environmental or social factors. Some models forecast the number of COVID-19 cases according to the population size \cite{kermack1927contribution,petropoulos2020forecasting,qin2020prediction}. Other models estimate the impact of the use of public transit and others study the contagion by airborne transmission \cite{gosce2018analysing, mo2020modeling}.

Our analysis leverages on the reported number of deaths for an accurate correlation because the reported number of cases are known to have large errors caused by the lack of testing and possible asymptomatic cases. 

In this analysis, we employ long short-term memory (LSTM) neural network for analyzing time series data to: 1) show the correlation between NYC subway turnstile (entry) data and COVID-19 deaths and cases, and Auto-Regressive Integrated Moving Average (ARIMA) to both 2) predict the dates when COVID-19 deaths would approach zero, according to the reported data, and 3) estimate the time when the contagion started in NYC by identifying the dates when the first cases and the first deaths occurred through analysis of the provided data. We compare the estimated dates with the reported dates. 

The analysis performed in this paper uses exclusively the mentioned public datasets, and other factors that are not included in the analyzed dataset may have affected the actual counts of cases and deaths. For example, the data analyzed in this paper was recorded while no face coverings were widely considered.

The remainder of this paper is organized as follows. Section \ref{sec:model} describes the models used in this paper for data analysis. Section \ref{sec:results} describes the obtained results. Section \ref{sec:discussion} provides a discussion and remaining questions. Section \ref{sec:conclusions} presents our conclusions.

\section{Model Description}
\label{sec:model}
In this section, we present the data sources and analysis tools used in this paper.

\subsection{Datasets}

The NYC DOH dataset reports the first COVID-19 case on February 29, 2020 and the first death on March 11, 2020. 
The MTA turnstile usage data \cite{MTA} includes the number of subway entries and exits per station in NYC, recorded hourly each day. The total number of stations in the dataset is 379. We calculate the average daily entries of NYC subway stations per day for our analysis. We also use the number of daily cases and deaths of the COVID-19 dataset published by the NYC DOH \cite{NYChealth}. 

We consider that there is an incubation and symptomatic period for people who lost their lives to COVID-19. We consider those periods as features to analyze the correlation of the studied data. We call these feature day-shifted subway entries. Shifted here means subway entries from days prior to the reported death's date.
This is, for a given day $t_0$, the turnstile usage is denoted as $x_{t_0}$ and the turnstile entry of $m$ days before $t_0$ is denoted as $x_{t_0-m}$. We use $x_{t_0}$ and $x_{t_0-m}$ as features to predict the number of deaths and cases for day $t_0$, where $1\leq m \leq 25$. We use 20\% of the NYC DOH data as test data for validation and the rest for training in our analysis. 

\subsection{Forecasting Models}
To analyze the turnstile usage data and its correlation to COVID-19 deaths and cases, we use linear regression and LSTM.  Linear regression is a statistical method for finding the relationship between a dependent variable and independent variables or features\cite{montgomery2012introduction}. In this paper, we use a linear regression model with L1 regularization to find the precedence of features. We adopt multiple linear regression with $x_j$ independent variables. The estimated value $\hat{y}$ in the linear regression model follows: 
\begin{equation}
  \hat{y} = \beta_0 + {{\sum\limits_j \beta_j x_{j}}} + \varepsilon 
\end{equation}
where ${\beta _j}$ represents the model weights or regression coefficients and $\varepsilon$ is the residuals error. L1 regularization is equal to the absolute sum of the model weights multiplied by a shrinkage value ($\lambda$). L1 regularization is formulated as $\lambda \sum\limits_j {|\beta_j|}$ and the regression model goal is to minimize:
\begin{equation}
  \sum\limits_i {({y_i} -\beta_0- {{\sum\limits_j \beta_j x_{ij}}})} ^2 + \lambda \sum\limits_j |\beta_j|
\end{equation}
where $y_i$ is the actual data point and $x_{ij}$ is the value of the independent variables for each data point.
\par

LSTM is a recurrent neural network (RNN) model \cite{hochreiter1997long} capable of learning both recent sequences of inputs and historical data. 
LSTM is mainly used to model time-series data  
to learn the time evolution of sequences of data \cite{ramsundar2018tensorflow}. We use a Savitzky–Golay filter to smooth out the prediction results from LSTM \cite{schafer2011savitzky}.
\par
The ARIMA model is used to forecast the time-series data \cite{box2011time}. ARIMA consists of three models: auto-regressive (AR), integration, and moving average (MA). It uses three hyper parameters: $p$, $d$, and $q$, where $p$ is the order of auto-regressive model, $d$ is the degree of difference, and $q$ is the order of moving average \cite{box2011time}. The ARIMA($p,d,q$) equation is defined as 
\begin{equation}
  \left( {1 - \sum\limits_{k = 1}^p {{\alpha _k}{L^k}} } \right){\left( {1 - L} \right)^d}{X_t} = \left( {1 + \sum\limits_{k = 1}^q {{\beta _k}{L^k}} } \right){\varepsilon _t}
\end{equation}
where $\alpha$ is the coefficient of discrete time linear equation of AR, $L$ is a time lag operator defined as $LX_t=X_{t-1}$, $X_t$ is the observed value at time $t$, $\beta$ is the coefficient for the noise term, $\varepsilon$, in MA \cite{arimaeq}.

\par
\begin{figure*}[!htbp]
    \centering
    \includegraphics[width=0.9\textwidth]{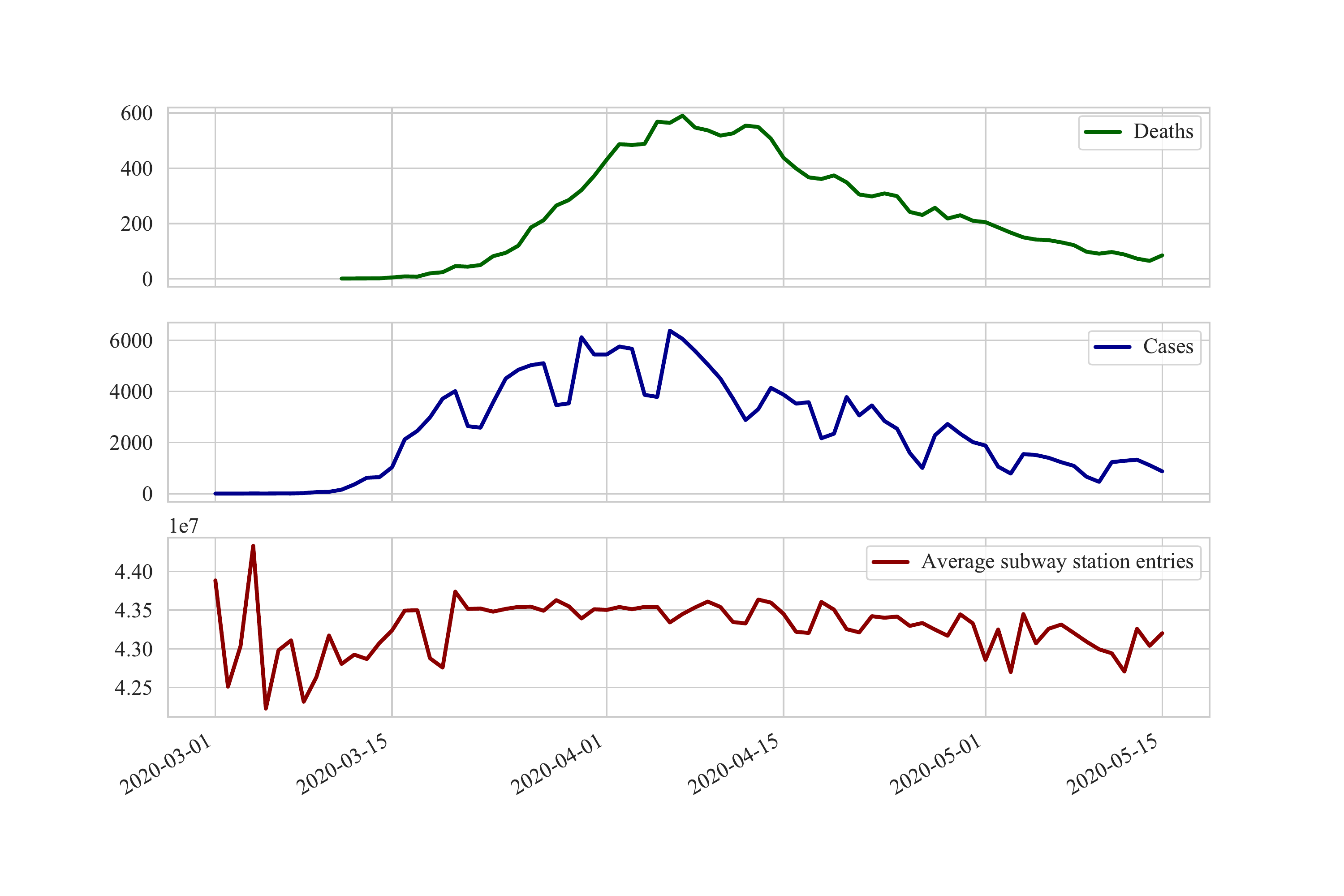}
    \caption{NYC deaths, cases, and MTA turnstile entries data from March 1, 2020 to May 15, 2020.}
    \label{fig:nyc_data}
\end{figure*}
\section{Results of Data Analysis}
\label{sec:results}

\subsection{Performance Metrics}
The performance of the prediction models is evaluated using $R^2$ score, mean absolute error (MAE), and root mean square error (RMSE). Coefficient of determination or $R^2$ score is the proportion of variation explained by independent variables \cite{montgomery2012introduction}. $R^2$ score is calculated as
\begin{equation}
    {R^2} = 1- \frac{{S{S_R}}}{{S{S_T}}}
\end{equation}
where ${S{S_R}}$ is the residual sum of squares and ${S{S_T}}$ is the total sum of squares. ${S{S_R}}$ is calculated as 
${\sum\limits_i {({y_i} - {{\hat y}_i})} ^2}$, where $y_i$ are the actual values and ${{\hat y}_i}$ are the predicted values. ${S{S_T}}$ is calculated as ${\sum\limits_i {({y_i} - \bar y)} ^2}$, where $y_i$ is the actual value and $\bar y$ is the mean of actual values. $R^2$ shows the variance of residuals (or prediction errors) in the predicted data points divided by the variance of data points from the average of all data points. $R^2$ represents the performance of a model as compared to randomly guessed predictions that are equal to the average of all data. The closer $R^2$ is to 1, the predictions are closer to the actual values.
\par
MAE is defined as the average of sum of residuals, or
\begin{equation}
    \frac{{\sum\limits_{i = 1}^n {\left| {{y_i} - {{\hat y}_i}} \right|} }}{n}
\end{equation}
where $n$ is the number of observations. When MAE is close to 0, the model has a high accuracy in predicting data points. We use MAE to measure the training and validation losses of LSTM. If training loss is significantly lower than validation loss, the model is overfitting the training data, which means the model is learning complex patterns of training data that may not generalize in predicting unseen test data and it results in poor performance. If validation loss is significantly lower than training loss, the model is underfitting the training data, which means that the model is unable to learn important patterns of the training data, and it results in poor performance of the model. As the training progresses if the losses grow apart, then we should stop the training and improve the model to avoid overfitting and underfitting \cite{tensorflow2015-whitepaper}.
\par
RMSE is defined as
\begin{equation}
    \sqrt {\frac{{\sum\limits_{i = 1}^n {{{\left( {{y_i} - {{\hat y}_i}} \right)}^2}} }}{n}}
\end{equation}
In other words, RMSE is the standard deviation of residuals. With normally distributed data points, one can expect 68\% of predicted points to be with one RMSE from the mean and 95\% to be within two RMSE from the mean of the actual data.

Figure \ref{fig:nyc_data} shows the average number of subway station entries and the reported number of COVID-19 deaths and cases each day starting from March 1, 2020 to May 15, 2020. The figure shows that the number of deaths peaked on April 7, 2020 before it started to decline. The number of cases peaked on April 6, 2020. The trend of the deaths curve trails the cases curve with a few days delay between March 15, 2020 to April 15, 2020.

\begin{figure}[H]
    \centering
    \includegraphics[width=\columnwidth]{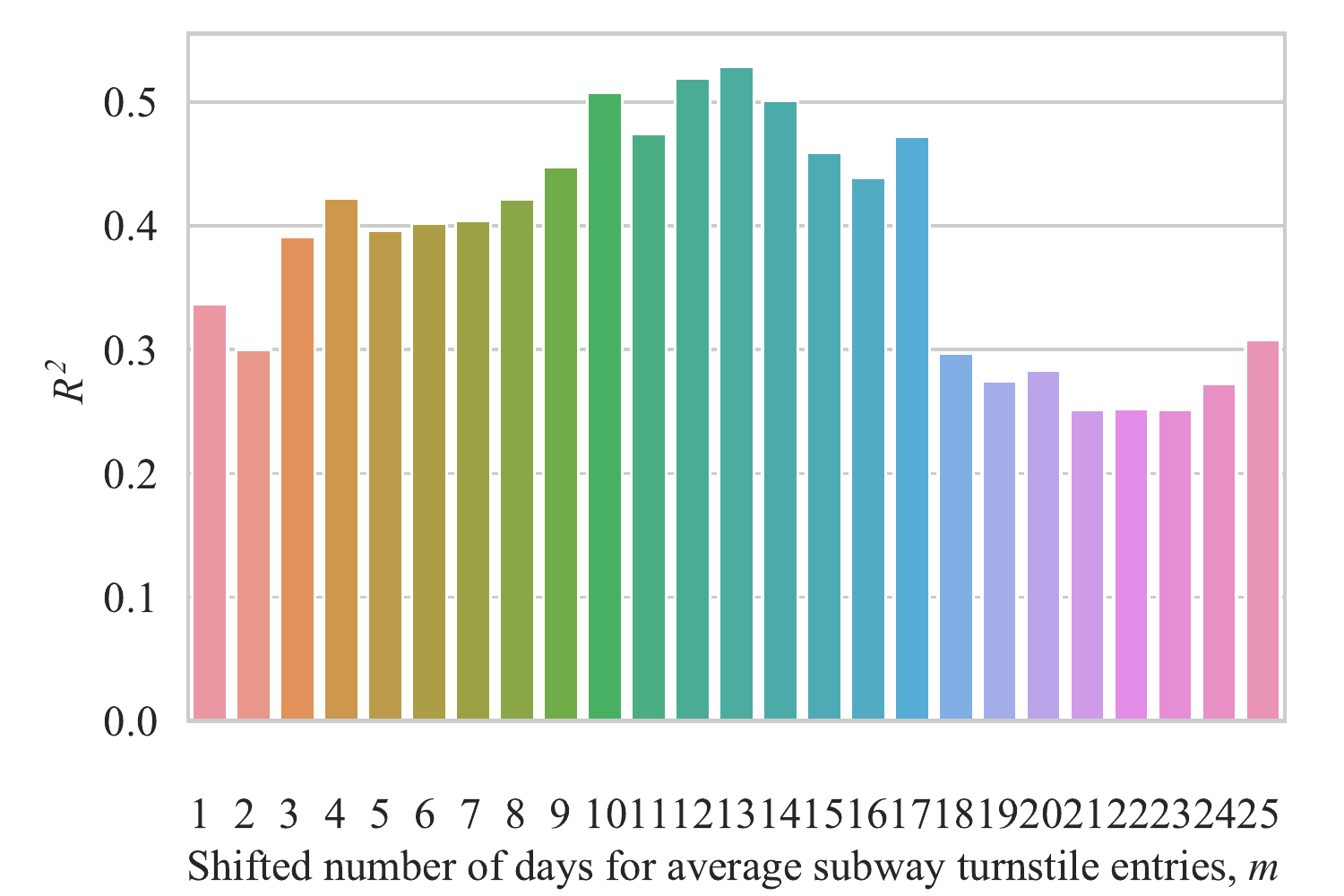}
    \caption{$R^2$ score for multiple regression prediction of COVID-19 deaths, with two independent variables: shifted average turnstile entries ($t_0-m$) and current average daily turnstile entries ($t_0$).}
    \label{fig:multi_linear_correlation}
\end{figure}

To find the correlation between NYC subway turnstile entries data and the number of reported COVID-19 deaths, 
we start from present day's turnstile usage data and then move it backwards $m$ days and use the historical data from the previous days to find the correlation between the number of COVID-19 deaths and subway entries. Figure \ref{fig:multi_linear_correlation} shows the calculated $R^2$ score in each test. 

The $x$-axis represents the independent variables of each test. For example, 1 shifted day (one day prior) in this figure shows the $R^2$ score obtained by the linear regression model for the turnstile usage of a single day and one day prior as a feature and the corresponding number of deaths reported on that day. The $y$-axis shows the $R^2$ score of the regression model between COVID-19 deaths on day $t_0$ and the subway entries on days $t_0$ and ${t_0-m}$. We calculated $R^2$ score for $0 \leq m \leq 25$, where we use the turnstile usage of prior 25 days.

From the $R^2$ scores shown in Figure \ref{fig:multi_linear_correlation}, we looked into those shifted days with high $R^2$ scores as candidates for features. We tested different combinations and the ones that produced the lowest MAE were selected. These selections are shown in Table \ref{tbl:featSets}. 
 
Feature set A consists of subway entries of the current day and 14 days prior (i.e., with one shifted-day data; that with the largest $R^2$). Feature set B consists of subway entries of the current day, 13, and 14 days prior (i.e., feature with two shifted-day data). Feature set C consists of subway entries of current day, 10, 12, 13, 14, and 17 days prior (i.e., feature with five shifted-day data, those with the highest $R^2$s). 
Feature set D adds two new features to Feature set C: the difference between subway entries of prior 8 and 13 days and that of prior 13 and 18 days (i.e., feature with two shifted days and difference from Friday to Monday for a weekly difference). We added these two new features with the speculation that changes in average subway ridership from Friday to Monday may also be indicators of trends. 
Feature set E includes shifted days with high $R^2$ but also shorter shifted days that have large $R^2$ (i.e., short and average incubation/symptomatic periods). 
Feature set F includes the combinations of the current date and prior-day data of up to 25 days (i.e., it includes all considered incubation/symptomatic periods in this paper that once they are added are 1 to 25 days). This set is selected for completeness and as a reference.

\begin{table*}[ht!]
\renewcommand{\arraystretch}{1.3} 
\centering
\caption{Feature Sets of Training Data Derived from Average Daily Entries of NYC Subway Stations.}
\label{tbl:featSets}
\begin{tabular}{@{}llc@{}}
\toprule

\textbf{Feature set A} &  & Current day and 14 days ago  \\ \midrule
\textbf{Feature set B} &  & Current day, 13, and 14 days ago \\ \midrule
\textbf{Feature set C} &  & Current day, 10, 12, 13, 14, and 17 days ago \\ \midrule
\textbf{Feature set D} &  & Current day, 10, 12, 13, 14, and 17 days ago, difference of 8 and 13 days ago, difference of 13 and 18 days ago \\ \midrule
\textbf{Feature set E} &  &  Current day, 4, 7, 11, 14, 15, and 17 days ago\\\midrule
\textbf{Feature set F} &  & Current day, 1, 2, 3, 4, 5,..., and 25 days ago \\
\bottomrule
\end{tabular}
\end{table*}

\subsection{Correlation between NYC Subway Usage and the Number of COVID-19 Deaths}

\begin{figure*}[htb!]
\centering
\subfigure[]{
\includegraphics[width=0.9\columnwidth]{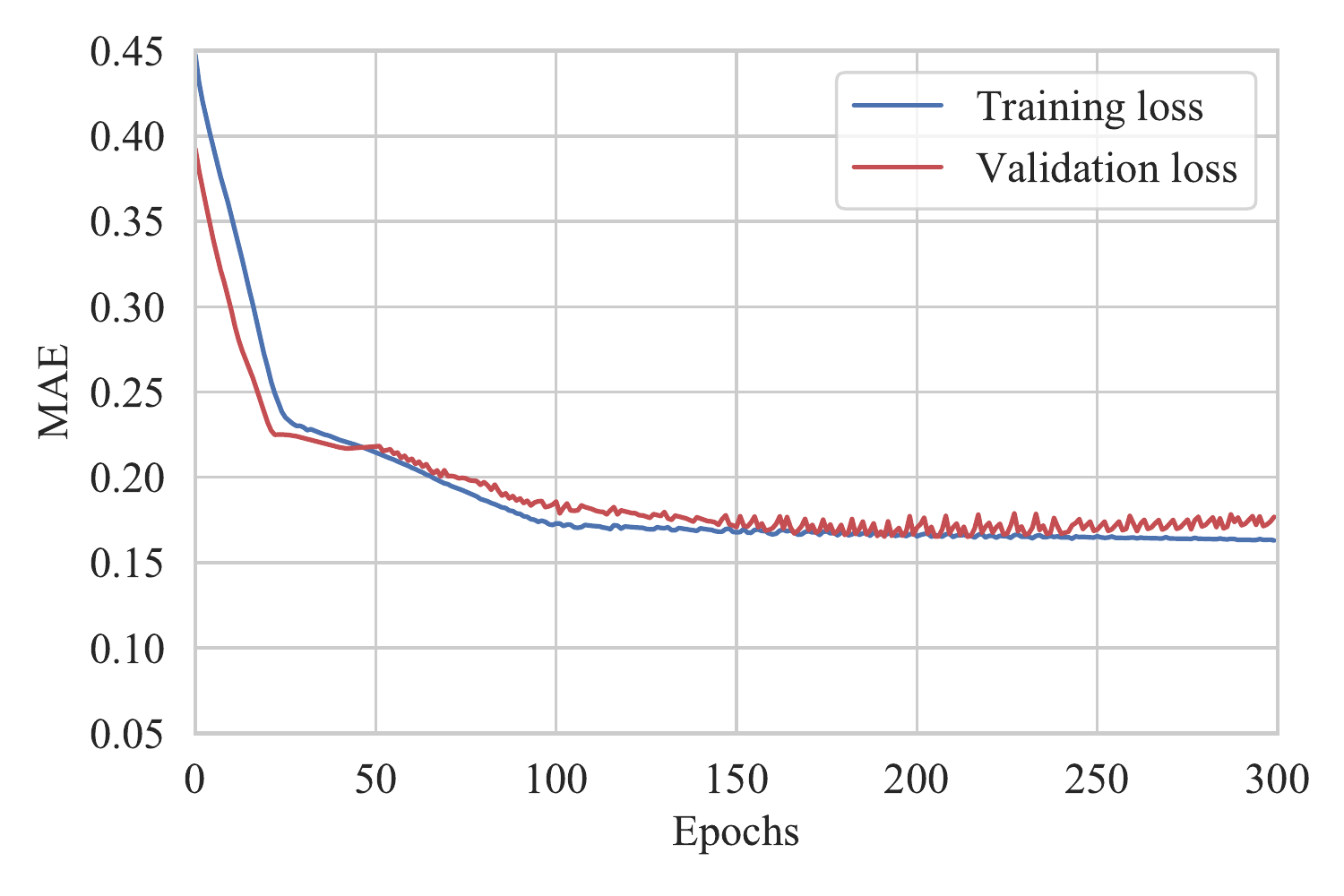}
\label{fig:a_mae}}
\subfigure[]{
\includegraphics[width=0.9\columnwidth]{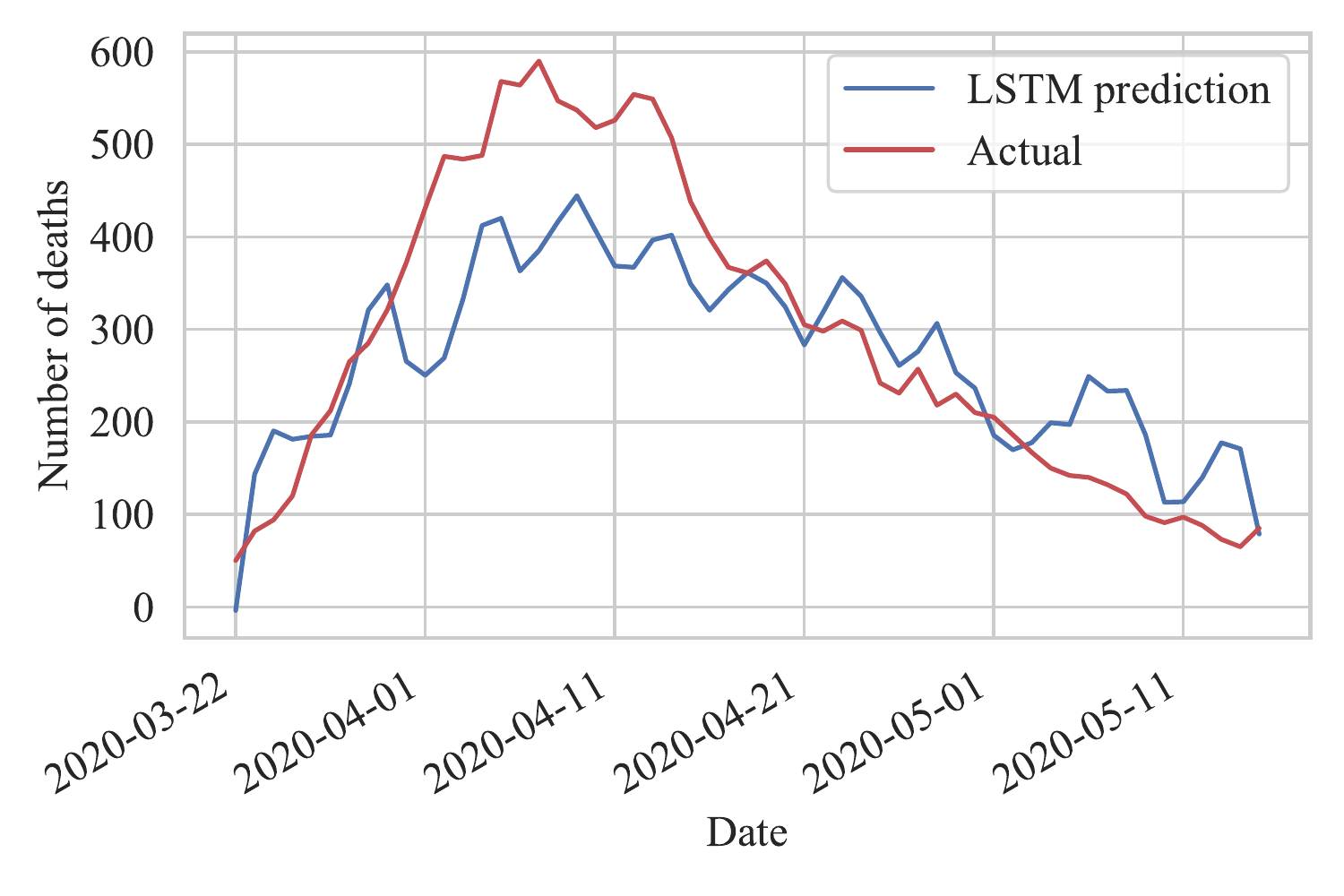}
\label{fig:a_deaths}}
\subfigure[]{
\includegraphics[width=0.9\columnwidth]{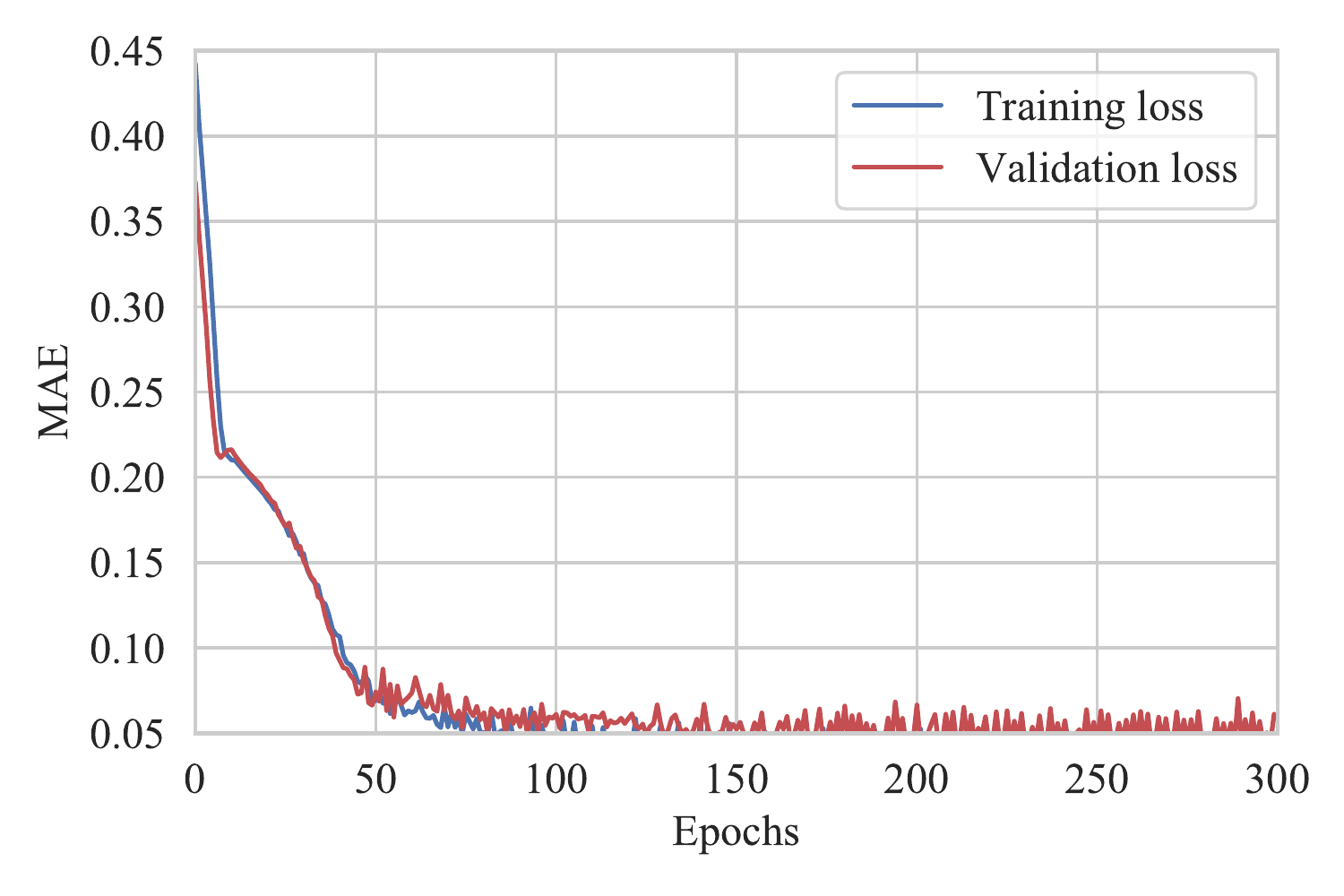}
\label{fig:f_mae}}
\subfigure[]{
\includegraphics[width=0.9\columnwidth]{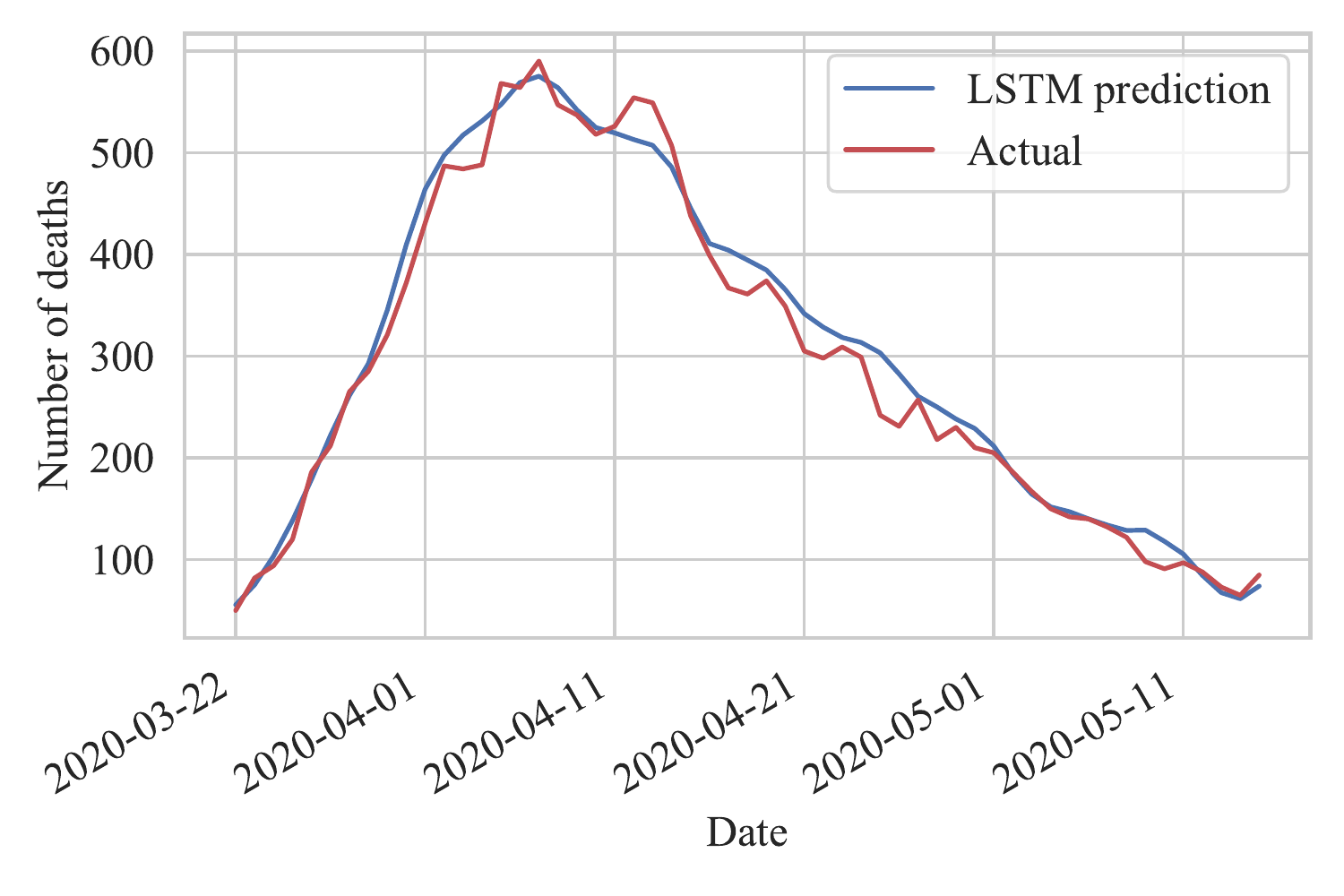}
\label{fig:f_deaths}}

\caption{Results with Feature set A of (a) MAE of LSTM training and validation in prediction of the number of deaths and (b) predicted number of COVID-19 deaths as a function of NYC MTA turnstile usage data. $R^2$ score on test data is 0.36 and RMSE is 78.06;  and  results with Feature set F of (c) MAE of LSTM training and validation in prediction of the number of deaths and (d) predicted number of COVID-19 deaths as a function of NYC MTA turnstile usage data. $R^2$ score on test data is 0.96 and RMSE is 25.66.}
\label{fig:lstm-deaths}
\end{figure*}

We applied the feature sets described in Table \ref{tbl:featSets} to LSTM to predict the number of deaths and cases. We compared the predicted number of deaths and cases using different features with the actual reported data.  
Figures \ref{fig:a_mae} and \ref{fig:a_deaths} show the MAE and the prediction of the number of deaths, respectively, of the LSTM model with Feature set A. The $R^2$ score for Feature set A is 0.36 and the RMSE is 78.06. Figures \ref{fig:f_mae} and \ref{fig:f_deaths} show the results for Feature set F. These results show that Feature set F produces the most accurate prediction with an $R^2$ score of 0.96 and an RMSE of 25.66. These two features show the best and worst performance of the predictions, as shown in Table \ref{tbl:scores}. This table summarizes the $R^2$ scores and the RMSEs for Feature sets A - F for deaths and cases.



\begin{figure*}[htb!]
\centering
\subfigure[]{
\includegraphics[width=0.9\columnwidth]{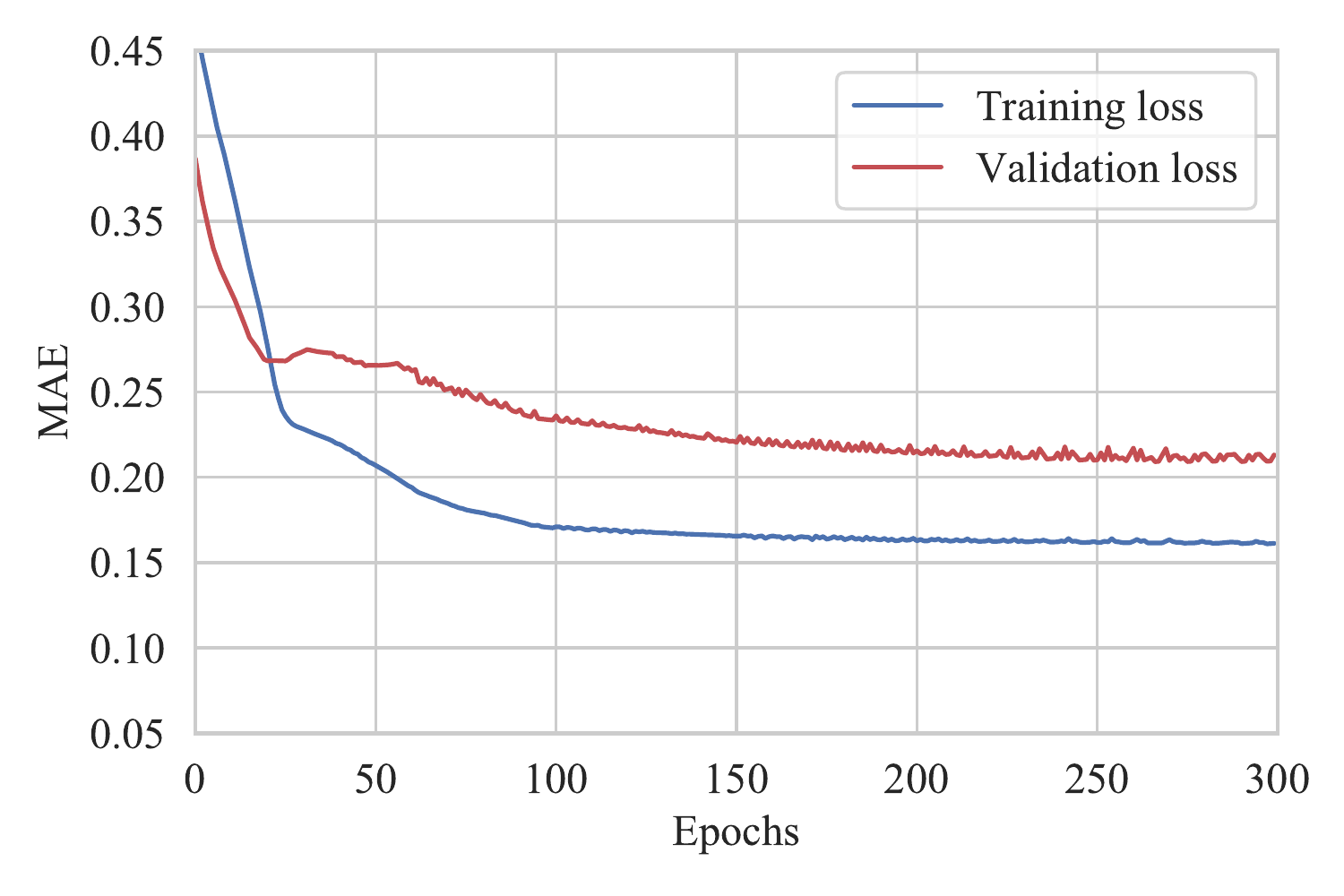}
\label{fig:a_cases_mae}}
\subfigure[]{
\includegraphics[width=0.9\columnwidth]{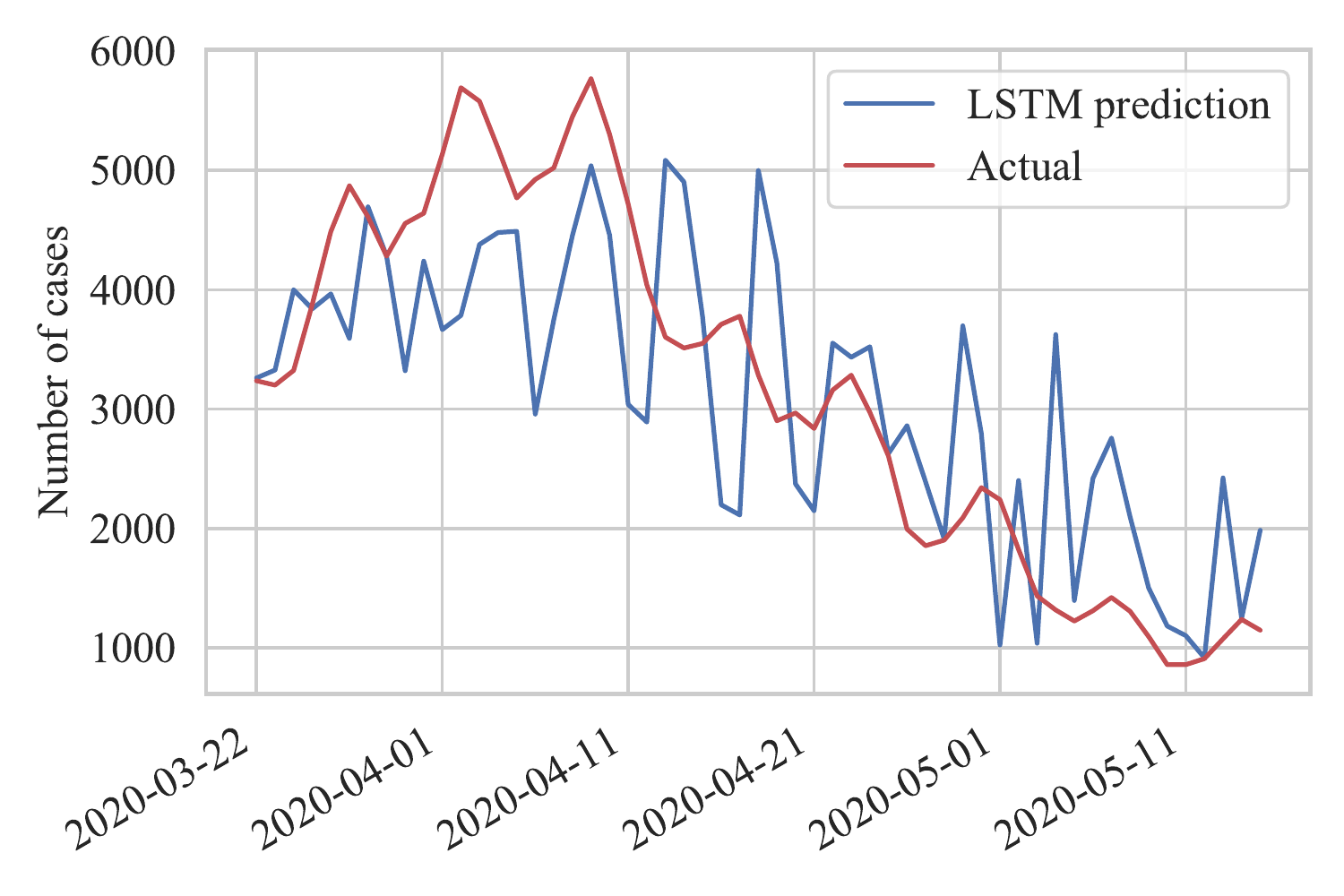}
\label{fig:a_cases}}
\subfigure[]{
\includegraphics[width=0.9\columnwidth]{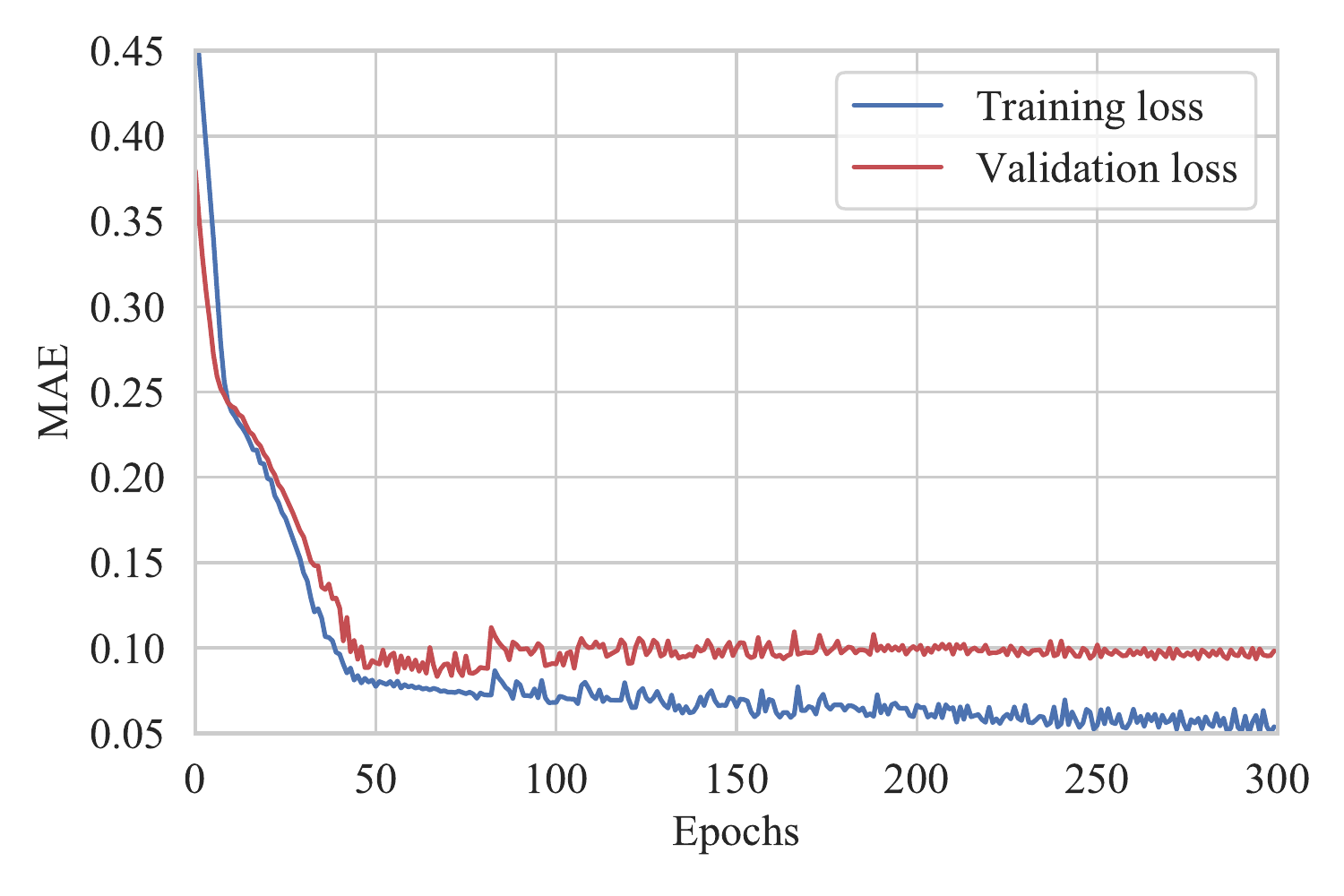}
\label{fig:f_cases_mae}}
\subfigure[]{
\includegraphics[width=0.9\columnwidth]{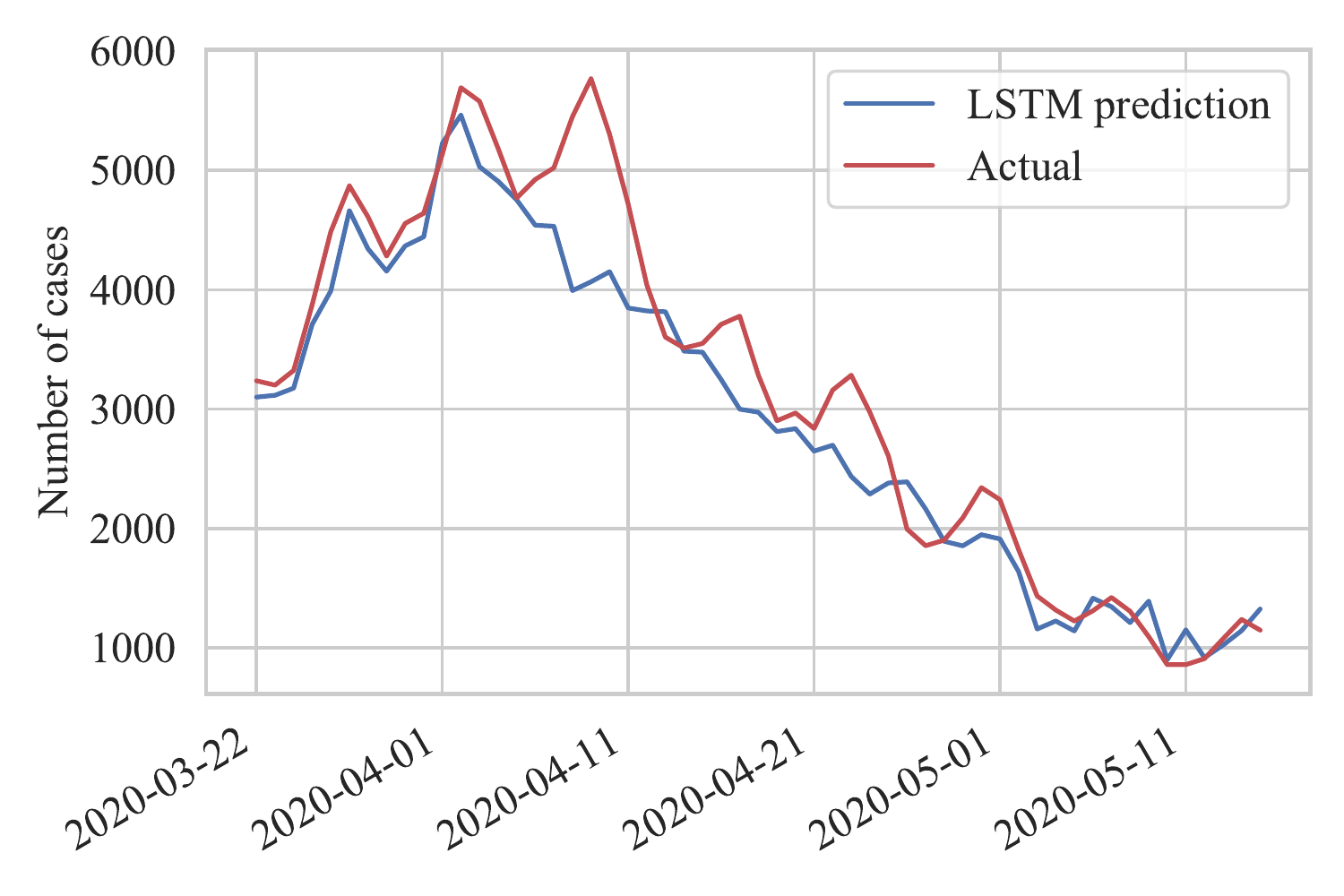}
\label{fig:f_cases}}
\caption{Results with Feature set A of (a) MAE of LSTM training and validation to predict the number of cases and (b) predicted number of COVID-19 cases as a function of NYC MTA turnstile usage data. $R^2$ score on test data is 0.49 and RMSE is 851.85; results with Feature set F of (c) MAE of LSTM training and validation in the prediction the number of cases and (d) predicted number of COVID-19 cases as a function of NYC MTA turnstile usage data. $R^2$ score on test data is 0.91 and RMSE is 414.61.}
\label{fig:lstm_cases}
\end{figure*}

We also used Feature sets A-F to predict the number of cases. Figure \ref{fig:a_cases_mae} shows the MAE and training and validation losses for feature set A. Here, the validation loss is slightly higher than the training loss. But both losses become steady after ~100 epochs. The LSTM predictions on Feature set A (Figure \ref{fig:a_cases}) follow the trend of the actual data. 
However, the predicted cases are less accurate than the predicted deaths because the reported cases are not accurate enough as compared to the reported deaths. The $R^2$ score on test data is 0.49 and the RMSE for all predicted data points is 851.85, which explains the observed discrepancies.
Figure \ref{fig:f_cases_mae} shows the training and validation losses of LSTM predicting number of cases by using Feature set F. Here, both losses are lower than those with Feature set A, which shows the high performance of LSTM in predicting cases. Additionally, the LSTM prediction in Figure \ref{fig:f_cases} follows the actual data closely. Table \ref{tbl:scores} also summarizes the $R^2$ scores and RMSE for each feature set. We observe a higher $R^2$ score does not necessarily lead to a more accurate prediction results as indicated in the cases prediction. The RMSEs for the case predictions are much higher than that of the deaths predictions. This further confirms the lack of accuracy in the reported cases.

\begin{table}[]
\renewcommand{\arraystretch}{1.3} 
\centering
\caption{RMSE of all data points and $R^2$ scores of test data for LSTM predicting deaths.}
\label{tbl:scores}
\begin{tabular}{@{}lll@{}}
\toprule
\textbf{Features - Deaths} &  $R^2$ & \textbf{RMSE}  \\ \midrule
\textbf{Set A}       &     0.12  &     78.06   \\
\textbf{Set B}       &     0.36  &     89.80   \\
\textbf{Set C}       &     0.49  &     70.52   \\ 
\textbf{Set D}       &     0.65  &     51.79   \\ 
\textbf{Set E}       &     0.69  &     54.93   \\ 
\textbf{Set F}       &     0.96  &     25.66   \\  \bottomrule
\textbf{Features - Cases}& &  \\ \cmidrule{0-0}
\textbf{Set A}    &       0.49 &   851.85           \\
\textbf{Set B}    &       0.30 &   1020.99          \\
\textbf{Set C}    &       0.45 &   729.5            \\
\textbf{Set D}    &       0.63 &   1043.5           \\
\textbf{Set E}    &       0.60 &   800.56           \\
\textbf{Set F}    &       0.91 &   414.61           \\\bottomrule
\end{tabular}
\end{table}

\subsection{Estimation of Trends and Dates of Minimum Cases and Deaths}
\begin{figure*}[htb]
\centering
\subfigure[]{
\includegraphics[width=1.0\columnwidth]{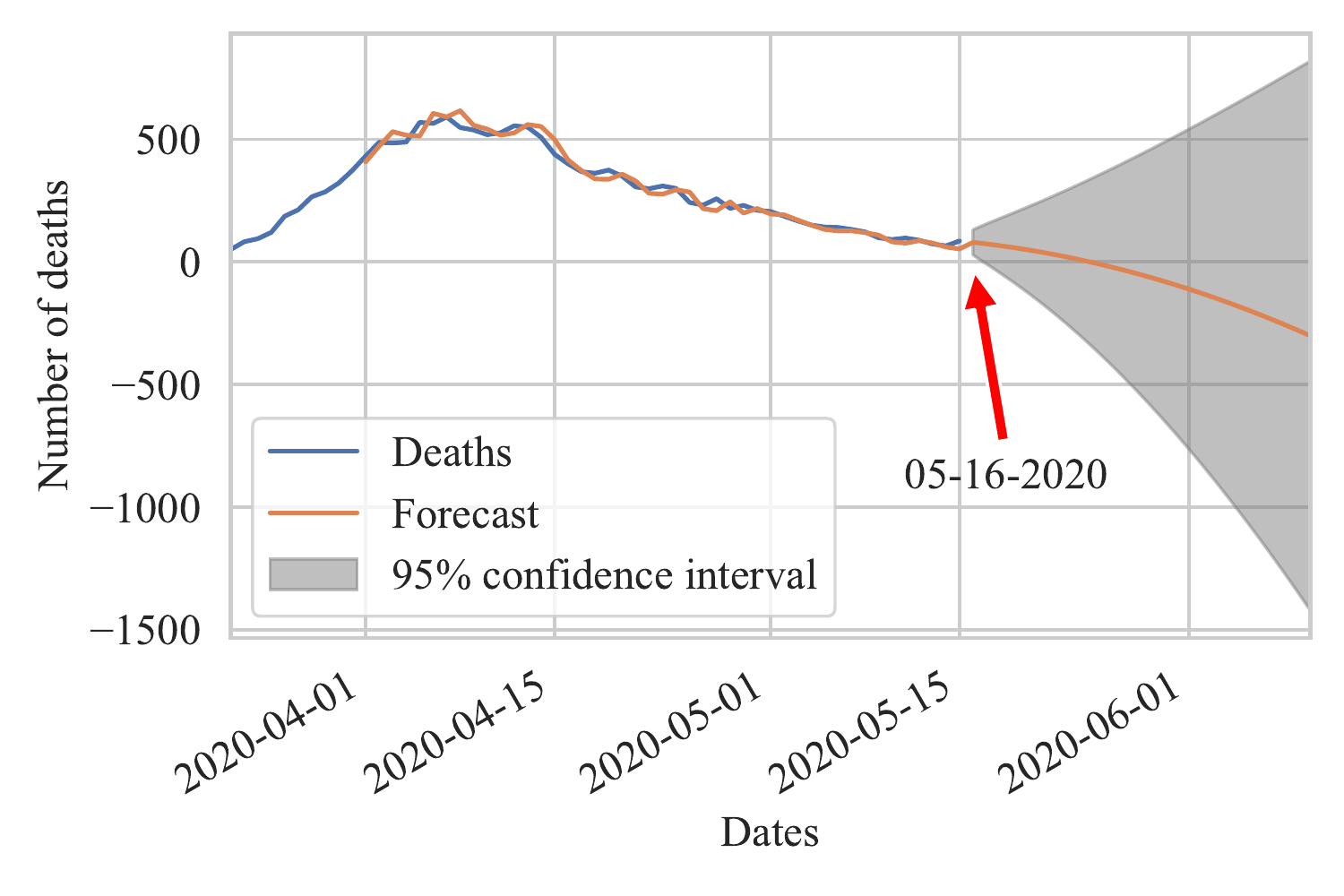}
\label{fig:arima_deaths}}
\subfigure[]{
\includegraphics[width=0.99\columnwidth]{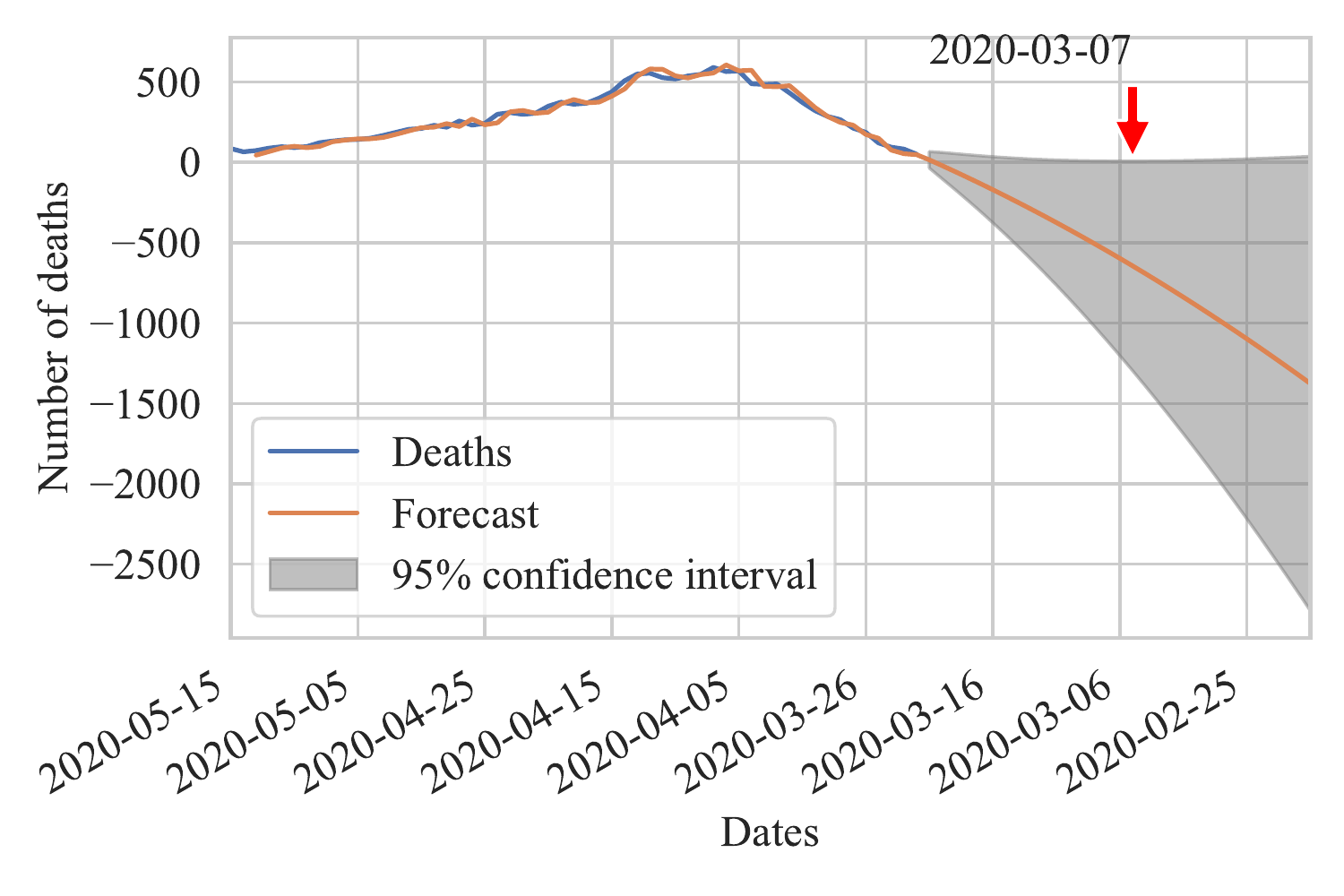}
\label{fig:arima_rev_deaths}}
\subfigure[]{
\includegraphics[width=1.0\columnwidth]{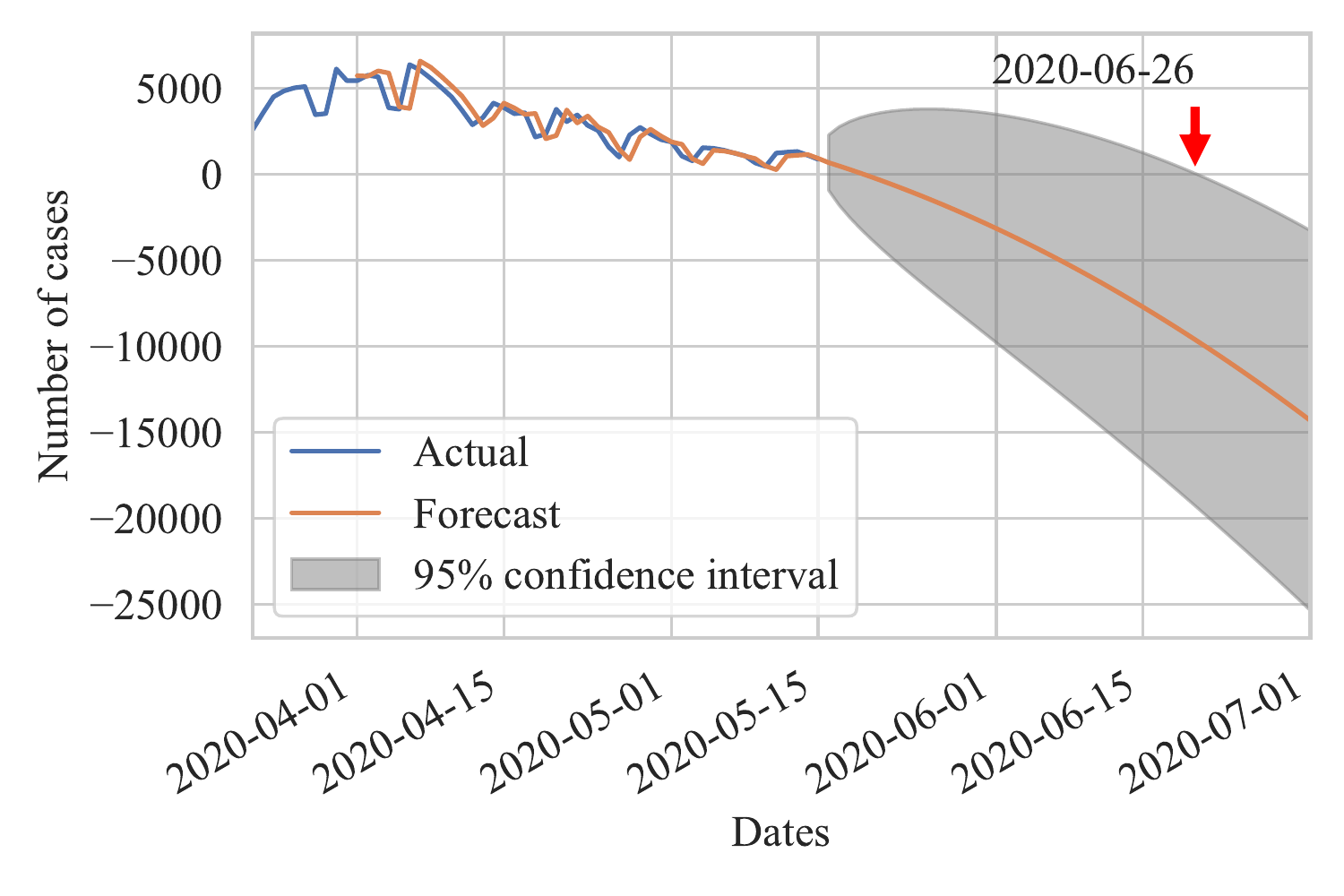}
\label{fig:arima_cases}}
\subfigure[]{
\includegraphics[width=1.0\columnwidth]{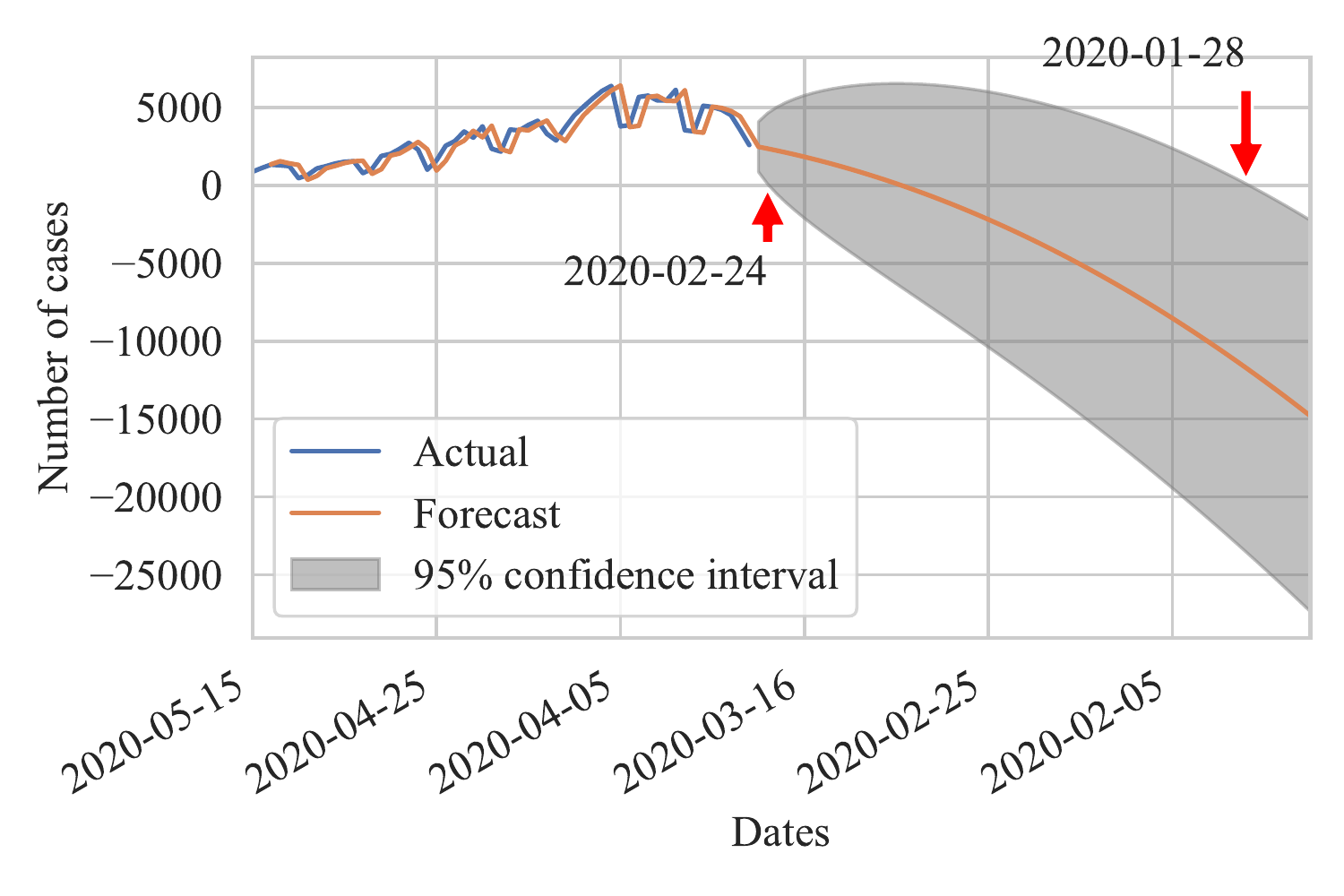}
\label{fig:arima_rev_cases}}
\caption{ARIMA forecast of (a) the number of COVID-19 deaths. The 95\% confidence interval of forecast after May 15, 2020 indicates that the number of deaths may increase or decrease. The lower bound of confidence interval marks May 16, 2020 as the day with zero deaths; (b) the number of deaths backward in time to find the possible first death. The upper bound of 95\% confidence interval of forecasts indicates Mar. 07, 2020 as the first possible death-case date; (c) the number of deaths backtracking in time to find the possible first death. The upper bound of 95\% confidence interval of forecasts indicates Mar. 07, 2020 as the first possible death-case date; (d) the number of cases in reverse chronological order to find the approximate dates of the occurrence of cases. The first case is projected to have occurred between January 28, 2020 and February 24, 2020.}
\label{fig:arima}
\end{figure*}

We use ARIMA analysis on the number of deaths to find the estimated date that the number of deaths in NYC reached zero. Figure \ref{fig:arima_deaths} shows that on May 16, 2020, the lower band of 95\% confidence interval of forecasts reaches zero. However, the upper bound does not cross zero value. That trend seems to indicate that the number of deaths could also grow again.

Figure \ref{fig:arima_cases} shows the ARIMA forecast of the progression of the number of COVID-19 cases in NYC after May 15, 2020. The forecast follows the decreasing trend of the number of cases. The red arrow here shows the date that the upper band of 95\% confidence interval crosses zero.

To find origin date of the first case in NYC, we used ARIMA to forecast the trend of COVID-19 cases in reverse chronological order, as Figure \ref{fig:arima_rev_cases} shows. The lower and upper band of 95\% confidence interval of ARIMA forecast shows that the first case of COVID-19 in NYC may have occurred between January 28, 2020 and February 24.

Table \ref{tbl:firstReports} shows the first reported COVID-19 cases (Feb. 29, 2020) and deaths (Mar. 11, 2020) in NYC DOH dataset.

\begin{table}[h]
\renewcommand{\arraystretch}{1.2} 
\centering
\caption{First reported COVID-19 case and death in NYC DOH dataset and the projected dates with ARIMA model.}

\begin{tabular}{@{}lll@{}}
\cmidrule[\heavyrulewidth]{1-3} 
\textbf{Date} & \textbf{Reported} & \textbf{Projected} \\ \cmidrule[\heavyrulewidth]{1-3} 
\textbf{First case}    &     Feb. 29, 2020 & Jan. 28, 2020 - Feb. 24, 2020\\ 
\textbf{First death}    &    Mar. 11, 2020 &  Mar. 7, 2020 - Mar. 21, 2020\\
\textbf{Date with 0 deaths}    &   Jul. 28, 2020 & May 16, 2020 - \_\_\_ \\ 
\textbf{Date with 0 cases}     & \_\_\_  & \_\_\_ - Jun. 26, 2020\\ 
\bottomrule
\end{tabular}
\label{tbl:firstReports}
\end{table}

\section{Discussion}
\label{sec:discussion}

The reporting of cases has been affected by the deployment and execution of testing for SARS-CoV-2. The effective and widespread testing offers better identification of the cases and their locations. But because of the limitation of testing capabilities at the beginning of the pandemic, the reported COVID-19 cases may not be accurate enough for modeling and thus resulting in larger errors. 
Therefore, the reported deaths data have been used to forecast and estimate the correlation between subway usage and the reported numbers of cases and deaths in this paper.

The estimations performed with ARIMA are based on the provided data with no additional information. The discrepancies on the estimated dates and the reported dates are the product of the model using solely past data.

Public health suggestions on the use of face covering and social distancing are not considered in the models to reflect the multi-variant effects in linear regression. The use of face covering at some point can lead to different results. Comparisons of NYC data with those of other cities that experienced an early face-covering measure is of interest for forecasting future prevalence or containment of the spread of this virus.

\section{Conclusions}
\label{sec:conclusions}

We used data on the NYC subway turnstile usage published by the Metropolitan Transportation Authority of New York City during the heavy prevalence of COVID-19 in the city and the data of confirmed cases and deaths reported by NYC Department of Health to investigate their correlation. 


Here, we show that by considering different incubation-symptomatic periods for subway users after traveling by subway as features, there is a strong correlation between the reported number of COVID-19 deaths and the number of NYC subway passengers through long short-term memory analysis for the first time.
We have also shown that the ARIMA model can predict the dates when the numbers of deaths and cases are reduced to zero and estimate the possible dates of when the first case and death occurred at the beginning of the pandemic in NYC. This analysis is performed by using the reported cases and deaths data. 

We have shown the discrepancy of the estimated and reported dates in such scenarios and showed that some of these dates are close. In these models, we considered that the recorded number of cases may have large errors as testing for COVID-19 was not widely available nor easy to apply at the beginning and the peak of the pandemic in the first half of 2020.\\ 

\par
\noindent\textbf{Disclaimer:}
 Any opinions, findings, and conclusions or recommendations expressed in this material are those of the authors and do not necessarily reflect the views of their employers.






%


\bibliographystyle{IEEEtran}
\bibliography{mta_covid}

\end{document}